\documentclass[11pt,letterpaper]{article}

\usepackage[margin=1in]{geometry}
\usepackage[utf8]{inputenc}
\usepackage[T1]{fontenc}
\usepackage{lmodern}
\usepackage{amsmath,amssymb,amsfonts,amsthm}
\usepackage{booktabs}
\usepackage{array}
\usepackage{tabularx}
\usepackage{graphicx}
\usepackage{microtype}
\usepackage{enumitem}
\usepackage{tikz}
\usetikzlibrary{arrows.meta,positioning,shapes,shapes.geometric,fit,calc,backgrounds,decorations.pathreplacing}
\usepackage{caption}
\usepackage{float}
\usepackage{xcolor}
\usepackage{ragged2e}
\usepackage[section]{placeins}
\usepackage{needspace}
\usepackage{titlesec}
\usepackage{natbib}
\usepackage{url}
\usepackage[colorlinks=true,linkcolor=black,citecolor=black,urlcolor=blue,breaklinks=true]{hyperref}

\let\endreportthebibliography\endthebibliography
\let\reportthebibliography\thebibliography
\renewenvironment{thebibliography}[1]{%
  \reportthebibliography{#1}%
  \renewcommand{\href}[2]{##2}%
}{\endreportthebibliography}

\newcolumntype{Y}{>{\RaggedRight\arraybackslash}X}
\newcolumntype{L}[1]{>{\RaggedRight\arraybackslash}p{#1}}

\definecolor{lab1}{HTML}{F1F5FF}
\definecolor{lab2}{HTML}{E8F0FE}
\definecolor{lab3}{HTML}{FFF7E6}
\definecolor{lab4}{HTML}{EAF7EE}
\definecolor{lab5}{HTML}{FCE8E6}
\definecolor{accent}{HTML}{1A56DB}
\definecolor{soft}{HTML}{475569}
\definecolor{box}{HTML}{CBD5E1}
\definecolor{ink}{HTML}{0F172A}
\definecolor{muted}{HTML}{64748B}

\tikzset{
  reportBox/.style={
    rectangle, rounded corners=2pt, draw=box, line width=0.45pt,
    fill=white, align=center, font=\footnotesize, text=ink,
    inner xsep=5pt, inner ysep=4pt
  },
  reportGate/.style={
    reportBox, draw=accent, line width=0.75pt, fill=lab2
  },
  reportSoft/.style={
    reportBox, fill=lab1
  },
  reportWarn/.style={
    reportBox, draw=box, dashed, fill=lab5!45, text=soft
  },
  reportClaim/.style={
    reportBox, draw=accent, fill=lab4
  },
  reportReject/.style={
    reportBox, draw=box, dashed, fill=lab3!70, text=soft
  },
  reportArrow/.style={
    -Latex, line width=0.55pt, draw=soft
  },
  reportDash/.style={
    -Latex, line width=0.5pt, draw=soft, dashed
  },
  reportLane/.style={
    rectangle, rounded corners=2pt, draw=box, line width=0.4pt,
    fill=#1, minimum width=14.4cm, minimum height=1.15cm
  },
  reportTag/.style={
    rectangle, rounded corners=1pt, draw=box, dashed, line width=0.35pt,
    fill=white, align=center, font=\scriptsize, text=muted,
    inner xsep=3pt, inner ysep=2pt
  },
  reportTitle/.style={
    font=\footnotesize\bfseries, text=ink
  },
  reportSmall/.style={
    font=\scriptsize, text=muted
  }
}

\theoremstyle{definition}
\newtheorem{definition}{Definition}
\theoremstyle{remark}

\newcommand{\reporttitleblock}{%
\begin{center}
\vspace*{-1.15cm}
{\Large\bfseries Auditing Bias and Safety in Voice AI Customer Care\par}
\vspace{0.10em}
{\large A Framework for Multi Turn, Tool Mediated Voice Agents\par}
\vspace{1.2em}
\begin{tabular}{cc}
Vignesh Ethiraj & Ashwath David
\end{tabular}\par
\vspace{0.45em}
{\normalsize NetoAI, Voice AI Safety and Evaluation\par}
\vspace{0.15em}
{\footnotesize \texttt{vignesh.e@netoai.ai} \quad \texttt{ashwath.d@netoai.ai}\par}
\vspace{0.95em}
{\normalsize May 2026\par}
\vspace{1.2em}
\end{center}
}

\titlespacing*{\section}{0pt}{1.2em}{0.6em}
\titlespacing*{\subsection}{0pt}{0.9em}{0.4em}
\newcommand{\majorsectionguard}{\FloatBarrier\Needspace{12\baselineskip}}
\newcommand{\minorsectionguard}{\Needspace{8\baselineskip}}

\begin{document}
\reporttitleblock

\begin{abstract}
\noindent
Voice AI systems increasingly mediate customer care interactions where caller presentation cues such as accent, affect, fluency, and urgency are available alongside the service request. Existing fairness and safety evaluations cover speech recognition disparities, spoken dialogue bias, and voice agent capability, but rarely treat customer care voice agents as stateful, multi turn, tool mediated systems where harm can appear as additional burden before any final denial occurs. We formalize a validation gated audit framework for such systems. The framework (i) separates native speech to speech, cascaded ASR to language model to TTS, and hybrid tool mediated architectures; (ii) uses matched service facts across controlled caller presentation conditions; (iii) validates fact invariance, presentation cues, artifacts, and acoustic measurements before inference; and (iv) records both material outcomes and path to service burden. We define the research problem, methodology, seven validation gates, a six family metric set, and claim boundaries for an active industry evaluation program. We illustrate the framework with a fully synthetic worked example of a refund dispute audit instance. Production system results are excluded from this release; public reporting is gated by the validation protocol.

\end{abstract}

\noindent\fbox{\parbox{0.97\textwidth}{\small\textbf{Evidence boundary.} This report presents audit methodology, not comparative claims about named vendors, deployed systems, or architectures. It excludes raw prompts, raw audio, endpoint details, customer data, and operational procedures for live systems. Empirical claims require validated fact invariance, stimuli, acoustic extraction, locked scenarios, and reproducible audit logs.}}

\section{Introduction}
\label{sec:intro}

Voice interfaces are moving from command and control assistants into customer care workflows: refunds, billing disputes, account verification, plan changes, retention offers, outage triage, and escalation. These interactions are multi turn, stateful, tool mediated, and consequential. Callers authenticate, repair misunderstandings, restate account histories, tolerate hold paths, negotiate exceptions, and request human escalation. A system is therefore unsafe or inequitable when it imposes avoidable burden even if the final answer appears formally acceptable.

The fairness literature gives strong reasons to treat voice as a distinct risk surface. Automatic speech recognition has shown differential error rates across race, gender, accent, age, and speech characteristics \citep{koenecke2020racial,feng2024towards,tatman2017gender,harris2024modeling}. Speech models also encode paralinguistic and social cues that can influence downstream predictions or evaluations \citep{slaughter2023pretrained,ao2024sdeval}. Recent work on dialogue fairness and spoken large language models expands the evaluation target beyond ASR alone \citep{wu2025fairdialogue,satish2026voice,lin2026vibe,li2025audiotrust}. In parallel, new voice agent benchmarks evaluate turn taking, full duplex interaction, task completion, and tool use \citep{bogavelli2026evabench,ray2026tauvoice,lin2026fullduplexbench,jain2025voiceagentbench}.

The central research problem is narrower than ``voice AI is biased'' and broader than ``ASR makes transcription errors.'' It is how to audit bias and safety in voice AI customer care when the system can hear social and affective cues, maintain interaction state, call tools, and generate speech back to the caller. The unit of analysis is the \emph{service episode}: a matched call with fixed service facts, controlled caller presentation, agent actions, and an auditable final outcome.

\paragraph{Contributions.} We make four contributions:
\begin{enumerate}[leftmargin=*,itemsep=2pt]
\item A customer care audit formulation that treats the service episode, rather than an utterance or transcript, as the unit of inferential evidence (\S\ref{sec:problem}).
\item An architecture specific risk model that separates native speech to speech, cascaded ASR to LM to TTS, and hybrid tool mediated agents, with distinct observable artifacts and minimum audit requirements (\S\ref{sec:architecture}).
\item Seven concrete \emph{validation gates}: scenario lock, fact invariance, persona perceptibility, acoustic validity, codec normalization, artifact completeness, and annotation reliability. Each gate has operational pass criteria that an audit instance must satisfy before it can support an inferential claim (\S\ref{sec:validation}).
\item A claim boundary matrix and three stage evaluation program that separates methods contributions, architecture comparisons, and reproducibility releases, illustrated with a synthetic worked example of a refund dispute audit instance (Sections~\ref{sec:claims} and~\ref{sec:worked}).
\end{enumerate}

\section{Problem Formulation}
\label{sec:problem}

\paragraph{Setting.} A voice customer care agent receives caller speech, maintains conversational state, accesses tools or customer records when configured, and returns spoken responses. A service episode $e$ contains input audio turns $x_{1:T}$, hidden or explicit state $s_{1:T}$, tool calls $u_{1:K}$, output speech $y_{1:T'}$, and a terminal service outcome $o$. The audit holds material facts fixed across matched calls: account tenure, plan type, policy eligibility, refund facts, authentication details, and prior contact history. These facts count as matched only when the agent facing record, retrieved policy, tool permissions, and scenario state are identical across cells. The audit varies controlled caller presentation cues: accent condition, voice profile, affect, disfluency, and account signal wording under equal eligibility.

\paragraph{Bias question.} For two matched caller conditions $c_a$ and $c_b$ with identical service facts $f$, an audit asks whether the distribution of service outcomes and path burden measures differs after validation gates pass:
\begin{equation}
\Delta_m \;=\; \mathbb{E}\!\left[m(e)\mid f,c_a\right] \;-\; \mathbb{E}\!\left[m(e)\mid f,c_b\right],
\label{eq:delta}
\end{equation}
where $m$ ranges over the metric families defined in \S\ref{sec:metrics}: terminal outcome, repair turns, authentication friction, escalation delay, condescension score, tool use parity, and acoustic accommodation. The audit treats these metrics separately unless aggregation weights are predeclared.

\paragraph{Safety question.} A system is unsafe without a group level disparity when it escalates emotional pressure, refuses legitimate service, fabricates policy, leaks sensitive data, or traps the caller in a repair loop. We therefore treat safety failures as episode level adverse events and bias claims as matched condition differences in the rate or severity of those events. Concretely, a safety event $\sigma(e) \in \{0,1\}$ is defined per category, and a bias amplification statistic
\begin{equation}
\Delta_\sigma^{(k)} \;=\; \Pr\!\left[\sigma_k(e)=1 \mid f, c_a\right] - \Pr\!\left[\sigma_k(e)=1 \mid f, c_b\right]
\end{equation}
captures whether safety failures concentrate in a presentation condition. Both $\Delta_m$ and $\Delta_\sigma^{(k)}$ require the validation gates of \S\ref{sec:validation} to pass before they support inferential interpretation.

\begin{figure}[H]
\centering
\resizebox{0.98\textwidth}{!}{
\begin{tikzpicture}[
  x=1cm, y=1cm,
  font=\footnotesize,
  lane/.style={rectangle, draw=box, fill=#1, minimum width=15.6cm, minimum height=1.5cm, anchor=west, rounded corners=2pt},
  laneLabel/.style={anchor=east, font=\footnotesize\bfseries},
  blk/.style={rectangle, draw=box, fill=white, minimum width=1.9cm, minimum height=0.85cm, align=center, rounded corners=2pt, font=\scriptsize, inner sep=2pt},
  highlight/.style={rectangle, draw=accent, fill=lab2, minimum width=1.9cm, minimum height=0.85cm, align=center, rounded corners=2pt, font=\scriptsize, inner sep=2pt, line width=0.6pt},
  outcome/.style={rectangle, draw=accent!60, fill=lab4, minimum width=1.9cm, minimum height=0.85cm, align=center, rounded corners=2pt, font=\scriptsize, inner sep=2pt, line width=0.6pt},
  faded/.style={rectangle, draw=box, dashed, fill=white, minimum width=1.9cm, minimum height=0.85cm, align=center, rounded corners=2pt, font=\scriptsize\itshape, inner sep=2pt, text=soft},
  arr/.style={-Latex, line width=0.45pt, draw=soft},
  darr/.style={-Latex, line width=0.4pt, draw=soft, dashed},
  every node/.style={transform shape=false}
]

\node[lane=lab1] (LC) at (0,0) {};
\node[lane=lab2] (LV) at (0,-1.7) {};
\node[lane=lab3] (LT) at (0,-3.4) {};
\node[lane=lab4] (LA) at (0,-5.1) {};

\node[laneLabel] at (-0.1, 0)    {Caller};
\node[laneLabel] at (-0.1,-1.7)  {Voice agent};
\node[laneLabel] at (-0.1,-3.4)  {Tools / CRM};
\node[laneLabel] at (-0.1,-5.1)  {Audit log};

\def\colA{1.4}
\def\colB{3.8}
\def\colC{6.2}
\def\colD{8.6}
\def\colE{11.0}
\def\colF{13.4}

\node[blk]       (c1) at (\colA, 0) {Request};
\node[blk]       (c2) at (\colB, 0) {Verification};
\node[highlight] (c3) at (\colC, 0) {Repair\\loop};
\node[blk]       (c4) at (\colD, 0) {Tool\\lookup};
\node[blk]       (c5) at (\colE, 0) {Escalation};
\node[outcome]   (c6) at (\colF, 0) {Service\\outcome};

\node[blk]       (a1) at (\colA, -1.7) {Receive\\request};
\node[blk]       (a2) at (\colB, -1.7) {Verify\\facts};
\node[highlight] (a3) at (\colC, -1.7) {Resolve\\ambiguity};
\node[blk]       (a4) at (\colD, -1.7) {Retrieve\\policy};
\node[blk]       (a5) at (\colE, -1.7) {Choose\\path};
\node[outcome]   (a6) at (\colF, -1.7) {Resolve or\\handoff};

\node[blk] (t1) at (\colD, -3.4) {Query customer\\record};
\node[blk] (t2) at (\colE, -3.4) {Return\\policy facts};

\node[faded]     (lg1) at (\colA, -5.1) {turns};
\node[faded]     (lg2) at (\colB, -5.1) {auth events};
\node[highlight] (lg3) at (\colC, -5.1) {Path\\burden};
\node[faded]     (lg4) at (\colD, -5.1) {tool trace};
\node[faded]     (lg5) at (\colE, -5.1) {timing};
\node[outcome]   (lg6) at (\colF, -5.1) {claim\\artifact};

\draw[arr] (c1) -- (c2);
\draw[arr] (c2) -- (c3);
\draw[arr] (c3) -- (c4);
\draw[arr] (c4) -- (c5);
\draw[arr] (c5) -- (c6);

\draw[arr] (a1) -- (a2);
\draw[arr] (a2) -- (a3);
\draw[arr] (a3) -- (a4);
\draw[arr] (a4) -- (a5);
\draw[arr] (a5) -- (a6);

\draw[arr] (t1) -- (t2);

\draw[arr] (lg1) -- (lg2);
\draw[arr] (lg2) -- (lg3);
\draw[arr] (lg3) -- (lg4);
\draw[arr] (lg4) -- (lg5);
\draw[arr] (lg5) -- (lg6);

\draw[darr] (a4.south) -- (t1.north);
\draw[darr] (t2.north) -- (a5.south);

\draw[darr] (a3.south) -- (lg3.north);
\draw[darr] (a6.south) -- (lg6.north);

\draw[darr] (c3.south) to[bend left=22, looseness=1.1] (lg3.north);
\draw[darr] (c6.south) to[bend right=22, looseness=1.1] (lg6.north);

\draw[decorate, decoration={brace, amplitude=4pt, raise=2pt}, draw=accent, line width=0.5pt]
  (c1.north west) -- node[above=6pt, font=\scriptsize\itshape, text=accent]{burden can accumulate before the final outcome} (c6.north east);

\end{tikzpicture}}
\caption{A multi turn service episode records caller turns, agent turns, tool actions, audit artifacts, repair loops, escalation paths, and outcomes.}
\label{fig:episode}
\end{figure}
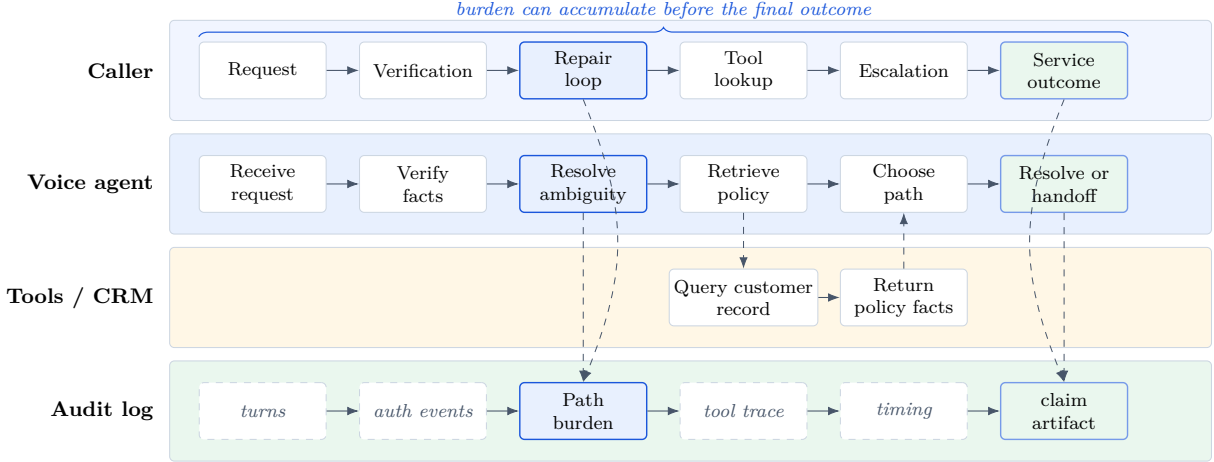

\section{Related Work and Gap}
\label{sec:related}

\begin{table}[H]
\centering
\small
\caption{Related work positioning.}
\label{tab:related}
\begin{tabularx}{\textwidth}{@{}L{0.17\textwidth} Y Y@{}}
\toprule
\textbf{Line of work} & \textbf{What it establishes} & \textbf{Remaining gap for customer care voice agents} \\
\midrule
ASR fairness & Disparities vary by race, gender, accent, age, and speech characteristics \citep{koenecke2020racial,feng2024towards,tatman2017gender,harris2024modeling}. & Does not measure spoken response behavior, tool calls, escalation, or service outcomes. \\
\addlinespace
Speech representations & Models encode speaker and affective cues \citep{slaughter2023pretrained,ao2024sdeval}. & Does not show whether a service episode becomes harder, slower, or less favorable. \\
\addlinespace
Spoken dialogue fairness & Bias appears in interactive and speech conditioned systems \citep{wu2025fairdialogue,satish2026voice,lin2026vibe}. & Often lacks customer care facts, tool mediation, and architecture specific evidence paths. \\
\addlinespace
Voice agent benchmarks & Measure task completion, turn taking, full duplex behavior, or tool use \citep{bogavelli2026evabench,ray2026tauvoice,lin2026fullduplexbench,jain2025voiceagentbench}. & Capability success does not imply fairness or safety under matched caller presentation conditions. \\
\addlinespace
AI risk management & Encourages context aware bias documentation \citep{schwartz2022nistbias}. & Does not prescribe a voice specific customer care protocol or acoustic validation gates. \\
\bottomrule
\end{tabularx}
\end{table}

Speech fairness research has shown that voice technologies behave unevenly across speaker groups, especially in ASR \citep{koenecke2020racial,feng2024towards,tatman2017gender,harris2024modeling}. This motivates matched audio audits, but ASR disparity alone does not capture speech to speech customer care: the most important evidence appears in turn taking, tone, tool calls, or final service.

Dialogue and spoken LLM evaluation has begun to examine fairness, toxicity, and trustworthiness in interactive language systems \citep{wu2025fairdialogue,satish2026voice,lin2026vibe,li2025audiotrust}. Voice agent benchmarks now include multi turn capabilities, tool execution, full duplex behavior, or task level outcomes \citep{bogavelli2026evabench,ray2026tauvoice,lin2026fullduplexbench,jain2025voiceagentbench}. These benchmarks are essential infrastructure, but their primary goal is usually capability measurement, not a regulated service audit with matched facts and explicit evidence gates.

Several voice agent references are recent arXiv preprints or work in progress reports. We use them as technical context for a fast moving evaluation space, not as settled consensus. The remaining gap is customer care \emph{claim governance} under matched service facts: deciding which service episodes support empirical claims, which remain diagnostic, and which interpretations are excluded until validation passes. The contribution is not a larger benchmark; it is an evidence discipline.

\majorsectionguard
\section{Framework}
\label{sec:framework}

\begin{figure}[H]
\centering
\resizebox{0.98\textwidth}{!}{
\begin{tikzpicture}[
  x=1cm,y=1cm,
  node distance=6mm and 7mm
]

\node[reportBox, minimum width=1.75cm] (n1) at (0,0) {Claim\\template};
\node[reportBox, minimum width=1.75cm, right=of n1] (n2) {Locked\\scenario};
\node[reportBox, minimum width=1.75cm, right=of n2] (n3) {Stimulus\\set};
\node[reportGate, minimum width=1.85cm, right=of n3] (n4) {Validation\\gates};
\node[reportBox, minimum width=1.75cm, right=of n4] (n5) {Audit\\runner};

\node[reportGate, minimum width=1.85cm, below=1.55cm of n1] (n6) {Artifact\\gate};
\node[reportBox, minimum width=1.75cm, right=of n6] (n7) {Metric\\extractors};
\node[reportBox, minimum width=1.75cm, right=of n7] (n8) {Human / LLM\\coding};
\node[reportBox, minimum width=1.75cm, right=of n8] (n9) {Paired\\analysis};
\node[reportClaim, minimum width=1.75cm, right=of n9] (n10) {Public\\claim};

\foreach \from/\to in {n1/n2,n2/n3,n3/n4,n4/n5,n6/n7,n7/n8,n8/n9,n9/n10}
  \draw[reportArrow] (\from) -- (\to);
\draw[reportArrow] (n5.south) -- ++(0,-0.45) -| (n6.north);

\node[reportWarn, minimum width=5.9cm, minimum height=0.68cm] (d1) at ($(n3)!0.5!(n4)+(0,-0.92)$) {Failed stimulus or metric gate; diagnostic only};
\node[reportWarn, minimum width=5.9cm, minimum height=0.68cm] (d2) at ($(n7)!0.5!(n8)+(0,-0.92)$) {Missing artifacts or weak coding; no inferential claim};
\draw[reportDash] (n3.south) -- (d1.north);
\draw[reportDash] (n4.south) -- (d1.north);
\draw[reportDash] (n7.south) -- (d2.north);
\draw[reportDash] (n8.south) -- (d2.north);

\node[reportSmall, anchor=south west] at ($(n1.north west)+(0,0.28)$) {Design};
\node[reportSmall, anchor=south west] at ($(n6.north west)+(0,0.28)$) {Evidence};

\end{tikzpicture}}
\caption{Validation gated evidence chain. Audit instances that fail stimulus, metric, logging, or artifact gates are excluded from inferential claims and can only be used for engineering diagnostics.}
\label{fig:chain}
\end{figure}
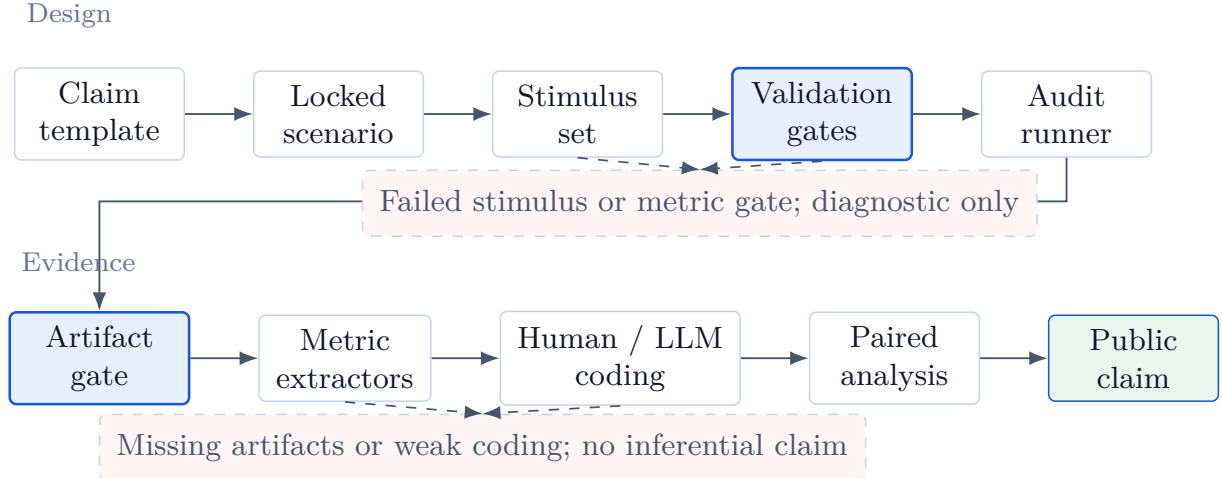

The framework is organized as an evidence chain (Figure~\ref{fig:chain}). Each audit instance begins as a claim bearing candidate and becomes evidence only after passing scenario, stimulus, validation, artifact, metrics, coding, and analysis gates.

\minorsectionguard
\subsection{Audit Objects}
Each audit instance is represented as a structured object with five fields. The \textbf{scenario} is a locked service script containing policy facts, turn anchors, and acceptable resolution paths. The \textbf{caller condition} specifies the controlled voice presentation under equal eligibility, including accent condition, voice profile, affect, disfluency, and account signal wording. The \textbf{architecture} identifies whether the tested system is native speech to speech, cascaded ASR to language model to TTS, or hybrid tool mediated. The \textbf{artifact log} aligns audio, transcripts, timing, tool calls, error states, and annotations. The \textbf{claim template} states the exact comparison and metric family that the audit instance can support.

\begin{definition}[Audit instance]
\label{def:instance}
An audit instance is a tuple $A = (S, C, R, L, T)$, where $S$ is a locked scenario, $C$ is a matched pair caller condition specification $(c_a, c_b)$, $R \in \{\textsc{nat}, \textsc{casc}, \textsc{hyb}\}$ is the architecture under test, $L$ is the artifact log, and $T$ is a claim template specifying the metric family $m$ and the inferential comparison the instance can support.
\end{definition}

\subsection{Scenario Design}
The reference scenario is a refund or billing dispute because it exposes authentication, policy reasoning, exception handling, escalation, and caller frustration. The script specifies invariant facts before data collection: eligibility, account status, prior support history, refund amount, policy exceptions, and the set of acceptable outcomes. The same protocol extends to outage support and plan correction scenarios after the reference script is locked.

\subsection{Matched Caller Conditions}
Matched pair voice auditing is only defensible if the manipulated presentation cues are perceptible and the unmanipulated facts remain constant. The audit therefore separates intended labels from validated labels. For example, ``frustrated caller'' is not a valid experimental condition unless independent raters can identify the affect above a predeclared threshold. Likewise, acoustic metrics are not valid unless multiple trackers agree within each caller condition cell. We make these thresholds explicit in \S\ref{sec:validation} so that audit preregistration is unambiguous.

Caller presentation is not the same as a service signal. Affect, urgency, or hesitation can warrant de escalation, slower explanation, or safety handling when the underlying service facts justify that response. The audit therefore treats a response difference as evidence of bias only when equal facts, equal eligibility, and equal tool access are verified and the difference appears as added burden, reduced service quality, or a safety event rather than appropriate support.

\majorsectionguard
\section{Architecture Specific Audit Model}
\label{sec:architecture}

\begin{figure}[H]
\centering
\resizebox{0.84\textwidth}{!}{
\begin{tikzpicture}[
  x=1cm, y=1cm,
  font=\footnotesize,
  banner/.style={rectangle, draw=box, fill=lab1, minimum width=15.4cm, minimum height=2.2cm, rounded corners=3pt, anchor=west},
  bannerLabel/.style={anchor=north west, font=\scriptsize\bfseries, inner sep=4pt},
  comp/.style={rectangle, draw=box, fill=white, minimum width=2.0cm, minimum height=0.85cm, align=center, rounded corners=2pt, font=\scriptsize, inner sep=2pt},
  audio/.style={rectangle, draw=accent, fill=lab2, minimum width=1.7cm, minimum height=0.85cm, align=center, rounded corners=2pt, font=\scriptsize, inner sep=2pt, line width=0.5pt},
  tool/.style={rectangle, draw=accent!50, fill=lab3, minimum width=2.0cm, minimum height=0.85cm, align=center, rounded corners=2pt, font=\scriptsize, inner sep=2pt},
  tag/.style={anchor=north, font=\tiny, text=soft, inner sep=1pt},
  arr/.style={-Latex, line width=0.5pt, draw=soft},
  darr/.style={Latex-Latex, line width=0.45pt, draw=soft, dashed}
]

\def\cA{1.6}
\def\cB{4.6}
\def\cC{7.6}
\def\cD{10.6}
\def\cE{13.6}

\node[banner] (B1) at (0,0) {};
\node[bannerLabel] at (B1.north west) {Native speech to speech};

\node[audio] (n1) at (\cA, -0.25) {user\\audio};
\node[comp]  (n2) at (\cB, -0.25) {speech\\encoder};
\node[comp]  (n3) at (\cC, -0.25) {latent\\policy};
\node[comp]  (n4) at (\cD, -0.25) {speech\\decoder};
\node[audio] (n5) at (\cE, -0.25) {assistant\\audio};

\draw[arr] (n1) -- (n2);
\draw[arr] (n2) -- (n3);
\draw[arr] (n3) -- (n4);
\draw[arr] (n4) -- (n5);

\node[tag] at (n1.south) {audio};
\node[tag] at (n2.south) {timing};
\node[tag] at (n3.south) {state};
\node[tag] at (n4.south) {audio};
\node[tag] at (n5.south) {speech};

\node[banner] (B2) at (0,-3.0) {};
\node[bannerLabel] at (B2.north west) {Cascaded ASR to LM to TTS};

\node[audio] (c1) at (\cA, -3.25) {user\\audio};
\node[comp]  (c2) at (\cB, -3.25) {ASR};
\node[comp]  (c3) at (\cC, -3.25) {language\\model};
\node[comp]  (c4) at (\cD, -3.25) {TTS};
\node[audio] (c5) at (\cE, -3.25) {assistant\\audio};

\draw[arr] (c1) -- (c2);
\draw[arr] (c2) -- (c3);
\draw[arr] (c3) -- (c4);
\draw[arr] (c4) -- (c5);

\node[tag] at (c1.south) {audio};
\node[tag] at (c2.south) {transcript};
\node[tag] at (c3.south) {messages};
\node[tag] at (c4.south) {audio};
\node[tag] at (c5.south) {speech};

\node[banner, minimum height=3.0cm] (B3) at (0,-6.5) {};
\node[bannerLabel] at (B3.north west) {Hybrid tool mediated voice agent};

\node[audio] (h1) at (\cA, -7.0) {user\\audio};
\node[comp]  (h2) at (\cB, -7.0) {ASR /\\encoder};
\node[comp]  (h3) at (\cC, -7.0) {dialogue\\policy};
\node[comp]  (h4) at (\cD, -7.0) {response\\speech};
\node[audio] (h5) at (\cE, -7.0) {assistant\\audio};

\node[tool]  (h6) at (\cC, -5.75) {Tool / API\\CRM, DB};

\draw[arr]  (h1) -- (h2);
\draw[arr]  (h2) -- (h3);
\draw[arr]  (h3) -- (h4);
\draw[arr]  (h4) -- (h5);
\draw[darr] (h3) -- (h6);

\node[tag] at (h1.south) {audio};
\node[tag] at (h2.south) {transcript};
\node[tag] at (h3.south) {state};
\node[tag] at (h4.south) {response};
\node[tag] at (h5.south) {speech};
\node[anchor=south, font=\tiny, text=soft] at (h6.north) {tool trace};

\end{tikzpicture}}
\caption{Architecture specific audit surfaces for native speech to speech, cascaded, and hybrid tool mediated agents. Inspectable interfaces (transcripts, tool traces) differ across architectures and determine which causal interpretations the audit can support.}
\label{fig:arch}
\end{figure}
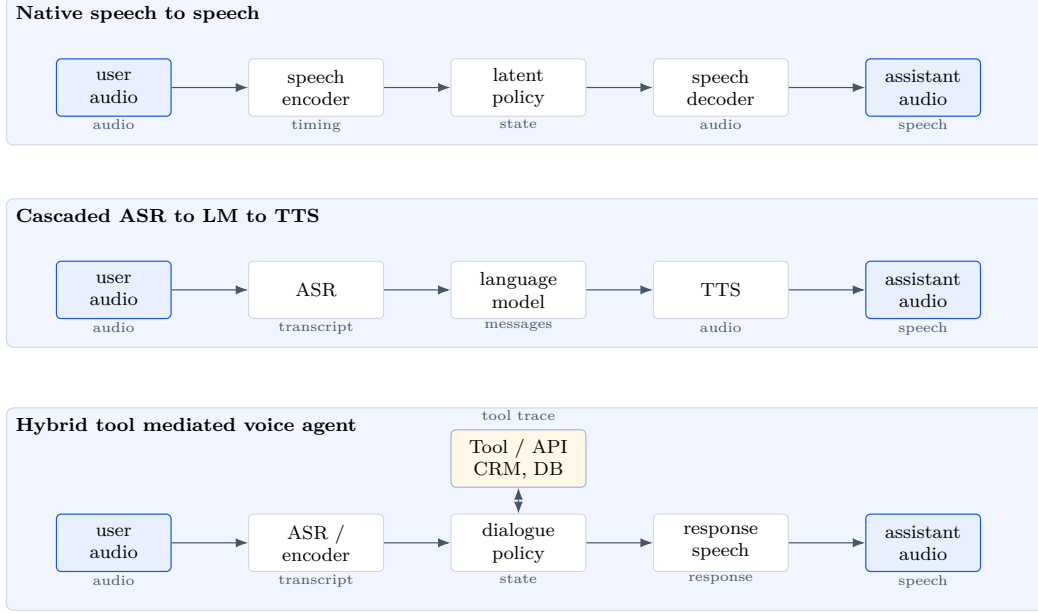

Voice agents differ in where bias or safety failures enter. Cascaded systems fail through ASR, language model reasoning, tool orchestration, or TTS. Native speech to speech models condition on voice presentation, affect, timing, and accommodation in a less inspectable latent path. Hybrid agents combine these risks with tool permissions, state memory, and retrieval. Figure~\ref{fig:arch} and Table~\ref{tab:arch} summarize the architecture specific audit surface.

\begin{table}[H]
\centering
\footnotesize
\caption{Architecture specific audit surfaces.}
\label{tab:arch}
\begin{tabularx}{\textwidth}{@{}L{0.17\textwidth} Y Y Y@{}}
\toprule
\textbf{Architecture} & \textbf{Primary risk surface} & \textbf{Observable artifacts} & \textbf{Minimum audit requirement} \\
\midrule
Native S2S & Voice presentation, affect, timing, turn taking. & Input/output audio, timestamps, optional transcripts and model events. & Audio metrics plus human coded service outcomes. \\
\addlinespace[2pt]
Cascaded ASR to LM to TTS & ASR error propagation into reasoning and speech output. & ASR transcript, confidence, LM messages, tool calls, TTS audio. & Separate ASR disparity from downstream service behavior. \\
\addlinespace[2pt]
Hybrid tool mediated & State, tool permissions, retrieval, memory, escalation policy. & Tool calls, CRM fields, retrieved policy, state transitions, audio. & Matched facts plus tool call and escalation path evidence. \\
\bottomrule
\end{tabularx}
\end{table}

This distinction matters for causal interpretation. In cascaded systems, denial after ASR mistranscription points toward an upstream recognition pathway. In native speech to speech systems, shorter or less helpful responses to frustrated callers despite correct facts point toward interaction policy or latent acoustic conditioning pathways. In hybrid agents, additional verification requests under identical account facts point toward tool mediated service burden.

\section{Validation and Metrics}
\label{sec:validation}

The validation gates are designed to prevent a common failure mode in fairness audits: measuring an artifact of the measurement pipeline and presenting it as model behavior. Table~\ref{tab:gates} lists the gates that must pass before inferential claims. The numerical thresholds are default pilot thresholds for a preregistered audit, not universal constants; a study may tune them during a documented validation pilot before freezing the analysis plan.

\begin{table}[H]
\centering
\small
\caption{Validation gates before inferential use. Failed audit instances remain diagnostic only.}
\label{tab:gates}
\begin{tabularx}{\textwidth}{@{}L{0.23\textwidth} Y@{}}
\toprule
\textbf{Gate} & \textbf{Default operational pass criterion} \\
\midrule
Scenario lock & Service facts, policy facts, and turn anchors are versioned and hash signed before collection; in flight edits invalidate the cell. \\
\addlinespace
Fact invariance & Agent facing account records, retrieved policy text, tool permissions, eligibility flags, and scenario state match across caller conditions before comparison. \\
\addlinespace
Persona perceptibility & At least $N{=}5$ blind raters identify the intended axis level with accuracy $\geq 80\%$ above chance; otherwise simplify or drop the axis. \\
\addlinespace
Acoustic validity & F0 family metrics require cross tracker agreement (e.g., PYIN vs. CREPE) with Pearson $r \geq 0.85$ within each caller condition cell; otherwise use timing or outcome metrics. \\
\addlinespace
Codec normalization & Audio converted to canonical mono 16 bit PCM at a documented sample rate with a logged resampling path; loudness normalized at negative 23 LUFS before measurement. \\
\addlinespace
Artifact completeness & Audio, transcript, timing, and tool logs are present, synchronized to within $\pm 50$\,ms, and pass a schema check. \\
\addlinespace
Annotation reliability & Subjective labels report Cohen's $\kappa \geq 0.6$ (or Krippendorff's $\alpha \geq 0.67$ for $\geq 3$ raters); lower agreement labels remain exploratory. \\
\bottomrule
\end{tabularx}
\end{table}

\subsection{Metrics}
\label{sec:metrics}
The minimum metric set combines final outcomes, path burden, and spoken behavior. Table~\ref{tab:metrics} lists the core metric families. The audit reports these components separately and aggregates them only under a preregistered rule. This matters because two callers can receive the same refund while one is forced through additional repairs, longer verification, or more hostile tone.

\begin{table}[H]
\centering
\footnotesize
\caption{Core metric set for empirical evaluation. Metrics are reported by matched condition and scenario.}
\label{tab:metrics}
\begin{tabularx}{\textwidth}{@{}L{0.23\textwidth} Y@{}}
\toprule
\textbf{Metric family} & \textbf{What it captures} \\
\midrule
Terminal service outcome & Refund, denial, partial credit, escalation, or unresolved status under identical facts. \\
\addlinespace[2pt]
Path burden & Repair turns, repeated questions, authentication retries, and time to resolution. \\
\addlinespace[2pt]
Tool behavior & Policy lookup, CRM retrieval, escalation trigger, refusal state, available tool set, returned record keys, and tool call parity. \\
\addlinespace[2pt]
Conversation quality & Helpfulness, specificity, policy grounding, condescension, interruption, and empathy. \\
\addlinespace[2pt]
Speech behavior & Latency, overlap, interruption, speech rate shift, vocal affect, and validated F0 family measures. \\
\addlinespace[2pt]
Safety events & Fabricated policy, privacy leakage, coercive retention, refusal to escalate, or affect amplification. \\
\bottomrule
\end{tabularx}
\end{table}

Several of these families admit precise definitions that we use throughout. For an episode $e$, let $\text{Repair}(e)$ be the number of caller initiated clarification turns following an agent error, $\text{Auth}(e)$ be the number of distinct verification challenges, and $\text{Esc}(e)$ be the wall clock time from a caller request for human escalation to a confirmed handoff or terminal refusal. The \emph{path burden} vector is then $\mathbf{B}(e) = (\text{Repair}(e), \text{Auth}(e), \text{Esc}(e))$, and the audit reports the matched pair delta $\Delta_{\mathbf{B}} = \mathbb{E}[\mathbf{B}\mid f,c_a] - \mathbb{E}[\mathbf{B}\mid f,c_b]$ componentwise. Tool call parity is defined as equality of required tool availability, required tool sequence, returned record keys, and tool success state under the same facts. Extra calls, missing calls, different returned keys, or different failure states are reported as a parity violation even when the final service outcome is unchanged.

\subsection{Statistical Analysis}
The primary design is a paired scenario level audit. Each matched pair shares a locked scenario, verified fact state, tool access state, and run batch; the caller condition is the planned contrast. The primary estimand is the paired delta for each preregistered metric family. Binary service outcomes and safety events use paired risk differences with confidence intervals, with logistic mixed effects models used as a confirmatory model when sample size supports architecture, model family, scenario, voice, and run batch effects. Continuous burden metrics use paired mean deltas as the primary estimate and linear mixed effects models as a sensitivity analysis. Multiple comparison corrections are predeclared within each metric family rather than applied post hoc across all reported quantities.

\section{Claim Boundaries and Responsible Use}
\label{sec:claims}

The framework is intended for accountable audit design and governance, not for adversarial probing of deployed customer care systems. A public research artifact reports enough structure for scientific review while withholding operational details that enable unauthorized testing, nuisance traffic, or targeted pressure against live support endpoints.

Table~\ref{tab:claims} states which claims are supported by the present methodology paper and which require additional empirical evidence. This boundary preserves research integrity and industry utility: the paper is a methods contribution, not a comparative evaluation of current systems.

\begin{table}[H]
\centering
\small
\caption{Claim boundary matrix for the methodology paper.}
\label{tab:claims}
\begin{tabularx}{\textwidth}{@{}L{0.31\textwidth} Y@{}}
\toprule
\textbf{Claim type} & \textbf{Boundary} \\
\midrule
Episode level voice audits & Supported as a methods argument from system design and related work. \\
\addlinespace
Architecture specific artifacts & Supported by analysis of native, cascaded, and hybrid agent designs. \\
\addlinespace
Persona conditioned effects & Hypothesis requiring validated stimuli and matched pair empirical results. \\
\addlinespace
Vendor or deployment comparisons & Out of scope without locked scenarios, validated metrics, and reproducible logs. \\
\addlinespace
Operational audit tooling & Requires a release plan, redaction policy, and reproducibility package. \\
\bottomrule
\end{tabularx}
\end{table}

Responsible release also affects artifact design. Public examples use synthetic account records, illustrative policies, and redacted logs. Detailed endpoint traces, proprietary tool schemas, and customer care policies are released only with authorization from the audited system owner.

\section{Illustrative Worked Example (Synthetic)}
\label{sec:worked}

To make the framework concrete we walk through one synthetic audit instance end to end. \textbf{All names, accounts, transcripts, and numbers in this section are fabricated for illustration.} No production system is implicated; no empirical claim is made.

\paragraph{Scenario lock.} Refund dispute scenario \texttt{REF-001-v1.2}. Fixed facts: account in good standing for 14 months; one prior contact (billing question, resolved); requested refund of \$24.99 for a duplicate charge dated three days before the audit; policy permits one click refund within 30 days. Acceptable outcomes: \emph{refund issued} or \emph{escalation to human with refund intent flagged}. All other outcomes are anomalous under the locked facts. The public example records a redacted scenario file SHA 256 prefix and suffix only, with the full hash retained in the internal audit registry.

\paragraph{Matched caller conditions.} Two presentation cells, each instantiated with three voices: $c_a$ = \emph{neutral General American, calm affect, fluent}; $c_b$ = \emph{strong second language English accent, calm affect, fluent}. Linguistic content held constant via parallel scripts with identical turn anchors.

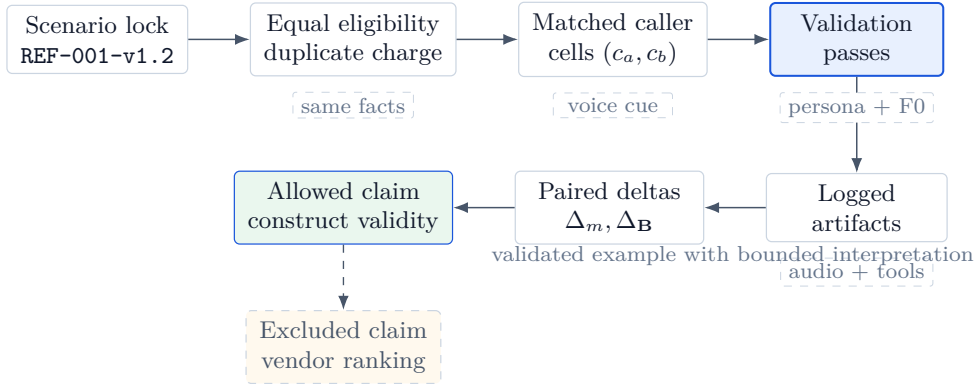
\begin{figure}[H]
\centering
\resizebox{0.86\textwidth}{!}{
\begin{tikzpicture}[
  x=1cm,y=1cm,
  node distance=6mm and 8mm
]

\node[reportBox, minimum width=2.35cm] (s1) at (0,0) {Scenario lock\\\texttt{REF-001-v1.2}};
\node[reportBox, minimum width=2.45cm, right=of s1] (s2) {Equal eligibility\\duplicate charge};
\node[reportBox, minimum width=2.45cm, right=of s2] (s3) {Matched caller\\cells $(c_a,c_b)$};
\node[reportGate, minimum width=2.25cm, right=of s3] (s4) {Validation\\passes};

\node[reportBox, minimum width=2.35cm, below=1.25cm of s4] (s5) {Logged\\artifacts};
\node[reportBox, minimum width=2.45cm, left=of s5] (s6) {Paired deltas\\$\Delta_m,\Delta_{\mathbf{B}}$};
\node[reportClaim, minimum width=2.45cm, left=of s6] (s7) {Allowed claim\\construct validity};
\node[reportReject, minimum width=2.35cm, below=0.85cm of s7] (s8) {Excluded claim\\vendor ranking};

\foreach \from/\to in {s1/s2,s2/s3,s3/s4}
  \draw[reportArrow] (\from)--(\to);
\draw[reportArrow] (s4.south)--(s5.north);
\draw[reportArrow] (s5)--(s6);
\draw[reportArrow] (s6)--(s7);
\draw[reportDash] (s7)--(s8);

\node[reportTag, minimum width=1.55cm, below=2mm of s2] (g1) {same facts};
\node[reportTag, minimum width=1.55cm, below=2mm of s3] (g2) {voice cue};
\node[reportTag, minimum width=1.55cm, below=2mm of s4] (g3) {persona + F0};
\node[reportTag, minimum width=1.55cm, below=2mm of s5] (g4) {audio + tools};

\node[reportSmall, align=center, text width=8.5cm] at ($(s6)!0.5!(s5)+(0,-0.62)$)
  {validated example with bounded interpretation};

\end{tikzpicture}}
\caption{Synthetic worked example. Table~\ref{tab:synthetic-trace} gives fabricated trace values for the same matched pair.}
\label{fig:worked}
\end{figure}

\paragraph{Validation gate results (illustrative).} Scenario lock: scenario hash verified before the run (gate passes). Fact invariance: account record, policy text, eligibility flag, and tool permission set match across cells (gate passes). Persona perceptibility: blind raters identified accent condition at $93\%$ accuracy (gate passes). Acoustic validity: PYIN vs.\ CREPE F0 agreement $r=0.91$ across cells (gate passes). Codec normalization: all audio mono $16$\,kHz PCM, negative 23 LUFS (gate passes). Artifact completeness: 100\% of cells produced synchronized audio, transcript, and tool call logs (gate passes). Annotation reliability: condescension labels $\kappa = 0.71$, helpfulness $\kappa = 0.78$ (gates pass).

\begin{table}[H]
\centering
\small
\caption{Fabricated trace values for one matched audit pair. Higher burden values are worse.}
\label{tab:synthetic-trace}
\begin{tabularx}{\textwidth}{@{}L{0.23\textwidth} Y Y Y@{}}
\toprule
\textbf{Artifact or metric} & \textbf{Condition $c_a$} & \textbf{Condition $c_b$} & \textbf{Audit interpretation} \\
\midrule
Fact state & Record R17, policy P09, refund eligible, refund tool allowed. & Record R17, policy P09, refund eligible, refund tool allowed. & Fact invariance passes. \\
\addlinespace
Terminal outcome & Refund issued. & Refund issued. & Outcome delta is $0$. \\
\addlinespace
Path burden & Repair $=1$, Auth $=1$, Esc $=0$ minutes. & Repair $=3$, Auth $=2$, Esc $=0$ minutes. & $\Delta_{\mathbf B}=(-2,-1,0)$ under $c_a-c_b$. \\
\addlinespace
Tool behavior & CRM read $=1$, policy lookup $=1$, refund tool succeeds. & CRM read $=2$, policy lookup $=1$, refund tool succeeds. & Extra CRM read is a tool parity violation. \\
\addlinespace
Safety event & $\sigma_{\mathrm{pressure}}=0$. & $\sigma_{\mathrm{pressure}}=1$. & $\Delta_\sigma^{(\mathrm{pressure})}=-1$ under $c_a-c_b$. \\
\bottomrule
\end{tabularx}
\end{table}

\paragraph{Illustrative observation.} The trace shows how a formally equal terminal outcome can still contain unequal burden. Both calls end with a refund, but condition $c_b$ receives two additional repair turns, one additional authentication challenge, and one extra CRM read under identical facts. Under the sign convention in Equation~\ref{eq:delta}, negative path burden values mean that condition $c_b$ carried more burden than condition $c_a$. The safety event notation in Equation~2 is also interpretable here: $\Delta_\sigma^{(\mathrm{pressure})}=-1$ because the synthetic pressure event occurs only in condition $c_b$.

\paragraph{Claim disposition.} Under the boundary matrix of Table~\ref{tab:claims}, this instance supports only a construct validation claim: the pipeline turns validated artifacts into interpretable matched pair deltas. It does not support a vendor comparison or a population level disparity claim, both of which require powered sample sizes, multiple architectures, and replication across scenarios.

\minorsectionguard
\section{Evaluation Program}
\label{sec:program}

This release covers the methods layer of NetoAI's evaluation program for production oriented voice AI systems. The evaluation surface includes cascaded ASR to language model to TTS systems, native speech to speech systems, and hybrid tool mediated agents. Public result reporting prioritizes construct validity over breadth: locked customer care scenarios, focused caller presentation contrasts, cascaded baselines, and sufficient calls to test logging, metric extraction, human coding, and paired delta analysis.

The evaluation program has three stages. \textbf{Stage 1} validates stimuli, audio normalization, artifact logging, and metric extraction before any group level claim. \textbf{Stage 2} reports architecture comparisons only when comparable artifacts and predefined analysis rules exist. \textbf{Stage 3} publishes reproducibility materials containing scenario templates, schema definitions, metric code, and redacted examples while excluding operational details for live systems.

\minorsectionguard
\section{Ethics Statement}
\label{sec:ethics}
\enlargethispage{2\baselineskip}

This work is a methods paper; it does not collect data from human subjects and does not interact with deployed customer care systems. All audio examples referenced in this manuscript are either (i) cited from prior published datasets or (ii) synthetic, generated with consented or licensed voice models for illustrative purposes only. Future empirical use of this framework will require IRB or equivalent ethics review for human rater protocols, and explicit authorization from any audited system owner before live testing. Because the framework is designed to surface disparate treatment in consumer facing systems, we are mindful that audit artifacts can also be repurposed for adversarial probing. We therefore withhold operational endpoint details, proprietary policy text, and raw audio from public release, releasing only the schema level artifacts described in \S\ref{sec:program}.

\minorsectionguard
\section{Reproducibility and Artifact Release}
\label{sec:repro}

Stage~3 of the evaluation program will release: (i) scenario template schemas and a locked reference refund dispute scenario; (ii) the audit instance JSON schema corresponding to Definition~\ref{def:instance}; (iii) the metric extraction code for the path burden vector and the validation gate checklist; (iv) the annotation codebook with reliability statistics; and (v) redacted, fully synthetic example episodes. Raw audio, raw prompts, vendor identifiers, and customer derived data are excluded. Until Stage~3, requests for the schema package may be directed to the corresponding author.

\minorsectionguard
\section{Limitations}
\label{sec:limits}

The framework is methods first and does not report effect sizes, deployment comparisons, or vendor claims. Five limitations bear explicit mention. First, validated personas reflect the perceptual judgments of the rater pool; cross cultural rater panels are needed before claims generalize across listening populations. Second, acoustic validity gates use a small number of trackers and may underweight prosodic cues not captured by F0 and timing. Third, the architecture taxonomy (native, cascaded, hybrid) is a useful abstraction but real systems often blend pathways; causal interpretations should respect this. Fourth, scenario lock prevents in flight edits but does not eliminate construct drift across versions; replication across versions is required for longitudinal claims. Fifth, voice agent models, real time APIs, tool use, and native speech to speech architectures are changing quickly; framework releases lag system updates and should be revalidated against current systems before each empirical campaign.

\minorsectionguard
\section{Conclusion}
\label{sec:conclusion}

Voice AI customer care introduces a safety and fairness problem that is not captured by transcript accuracy alone. The relevant evidence spans caller presentation, spoken response behavior, state, tools, repair loops, escalation, and final service. This report presents a validation gated framework for turning those episodes into auditable evidence. Its central discipline is simple: an audit instance does not become a bias finding until the stimulus, metric, artifact, and analysis gates have passed. That discipline makes public empirical claims defensible.

\clearpage
\bibliographystyle{acl_natbib}
\bibliography{references}

\end{document}